\documentclass[conference]{IEEEtran}
\IEEEoverridecommandlockouts
\usepackage{cite}
\usepackage{amsmath,amssymb,amsfonts}
\usepackage{algorithmic}
\usepackage{graphicx}
\usepackage{textcomp}
\usepackage{xcolor}
\def\BibTeX{{\rm B\kern-.05em{\sc i\kern-.025em b}\kern-.08em
    T\kern-.1667em\lower.7ex\hbox{E}\kern-.125emX}}
\usepackage{graphicx}
\usepackage{subcaption}
\graphicspath{ {images/} }
\usepackage[utf8]{inputenc}
\usepackage{fancyhdr}
\usepackage{dirtytalk}
\usepackage{tabularx}
\PassOptionsToPackage{hyphens}{url}\usepackage{hyperref}
\usepackage[ruled,lined,linesnumbered]{algorithm2e}

\usepackage{enumitem,kantlipsum}

\usepackage{graphicx}
\graphicspath{ {images/} }
\usepackage[utf8]{inputenc}
\usepackage{fancyhdr}
\usepackage{dirtytalk}
\usepackage{tikz}
\usetikzlibrary{shapes.geometric, arrows}
\usepackage{pgfplots}
\pgfplotsset{width=8cm,compat=1.16}
\usepgfplotslibrary{external}
\usepackage{mathtools}

\newtheorem{theorem}{\normalfont{\textbf{Theorem}}}

\tikzstyle{startstop} = [rectangle, rounded corners, minimum width=2cm, minimum height=0.5cm,text centered, draw=black, fill=red!30]
\tikzstyle{process} = [rectangle, minimum width=2cm, minimum height=0.5cm, text centered, draw=black, fill=orange!30, align=left]
\tikzstyle{decision} = [diamond, minimum width=1.0cm, minimum height=0.4cm, text centered, draw=black, fill=green!30]
\tikzstyle{arrow} = [thick,->,>=stealth]

\newcolumntype{L}{>{$}l<{$}}

\usepackage[labelformat=simple]{subcaption}

\begin{document}
\title{Semantic Information Market For The Metaverse: An Auction Based Approach }

\author{\IEEEauthorblockN{Lotfi Ismail\IEEEauthorrefmark{1},
Dusit Niyato\IEEEauthorrefmark{1},
Sumei Sun\IEEEauthorrefmark{2}, 
Dong In Kim\IEEEauthorrefmark{3}, 
Melike Erol-Kantarci\IEEEauthorrefmark{4}, and
Chunyan Miao\IEEEauthorrefmark{1}}
\IEEEauthorblockA{\IEEEauthorrefmark{1}School of Computer Science and Engineering,
Nanyang Technological University, Singapore}
\IEEEauthorblockA{\IEEEauthorrefmark{2}Institute for Infocomm Research, A*STAR, Singapore}
\IEEEauthorblockA{\IEEEauthorrefmark{3}Department of Electrical and Computer Engineering, Sungkyunkwan University, Suwon, South Korea}
\IEEEauthorblockA{\IEEEauthorrefmark{4}School of Electrical Engineering and Computer Science, University of Ottawa, Ontario, Canada
}
}


\maketitle
\begin{abstract}
In this paper, we address the networking and communications problems of creating a digital copy in the Metaverse digital twin. Specifically, a virtual service provider (VSP) which is responsible for creating and rendering the Metaverse, is required to use the data collected by IoT devices to create the virtual copy of the physical world. However, due to the huge volume of the collected data by IoT devices (e.g., images and videos) and the limited bandwidth, the VSP might become unable to retrieve all the required data from the physical world. Furthermore, the Metaverse needs fast replication (e.g., rendering) of the digital copy adding more restrictions on the data transmission delay. To solve the aforementioned challenges, we propose to equip the IoT devices with semantic information extraction algorithms to minimize the size of the transmitted data over the wireless channels. Since many IoT devices will be interested to sell their semantic information to the VSP, we propose a truthful reverse auction mechanism that helps the VSP select only IoT devices that can improve the quality of its virtual copy of objects through the semantic information.
We conduct extensive simulations on a dataset that contains synchronized camera and radar images, and show that our novel design enables a fast replication of the digital copy with high accuracy.
\end{abstract}


\begin{IEEEkeywords}
Metaverse, sensing-as-a-service, semantic communication, reverse auction.
\end{IEEEkeywords}


\section{Introduction}
\subsection{Background and Motivations}
Driven by the Covid-19 pandemic, the Metaverse has gained huge interest recently from different industry and public sectors~\cite{Jeong_2022, Papyshev2021ExploringCD, Minerva_2020}. Considered as the next generation of the Internet, the Metaverse enables users and objects to experience near real-life interaction with each other in the virtual environment through their avatars.
The Metaverse is made up from different emerging technologies such as virtual reality (VR), augmented reality (AR) and haptic sensors.
Furthermore, other emerging technologies such as beyond 5G and 6G are driving the Metaverse from imagination and fiction towards real world implementation as they enable users to access the Metaverse from anywhere, anytime instantly~\cite{Chen_2018}.

Digital twin modeling of the physical world in the Metaverse have a number of benefits for different application scenarios. For example, creating a virtual copy of the workplace enables employers to bridge the gap between working from office and working from home.
Another example appears in the smart vehicles area, where several companies face significant challenges to train their autonomous vehicles (AVs) in the real world.
A digital twin of the road, vehicles and other objects can be created, and then the AVs are trained in the Metaverse.
Instead of training the AVs on the real road network which is a high risk task, training in the Metaverse is much safer as road accidents in the Metaverse are not reflected to the real world, but the experiences are identical. More importantly, the experiences faced in the Metaverse are learnable by the learning model of the AVs~\cite{Amini_2020}.
An agriculture company can enable smart farming by creating a digital twin of the farm. IoT devices deployed in the field can collect several types of data about the plants and the soil, and then send it back to the edge server of the company through an unmanned aerial vehicle base station (UAV-BS) relay~\cite{Verdouw_2021_farming}. This enables farmers to simulate and observe effects of interventions with real-time data before physical interventions can take place.


As the digital twins are required to replicate the physical real-world system to the finest details~\cite{Minerva_2020}, generating an accurate 3D model of the physical system is the first step towards this goal.
However, the creation of an accurate 3D digital copy is challenging for several reasons. First, the collection of data about the environment requires different types of sensors, e.g., cameras, radars and LIDARs, which are required to be located at the region of interest of the VSP. It will be costly for the VSP to deploy IoT devices equipped with the aforementioned sensors for each task which are dynamic in both time and location. Second, the collected data by the IoT devices is huge in size and the available bandwidth for data transmission will quickly exceed the system limitation. 
To enable a real-time construction of the digital twin in the Metaverse, the communication system needs to be carefully designed as to maximize successful data transmission while minimizing the latency of packet delivery.

\subsection{Contribution}
In this paper, we study the paradigm of digital twin construction over wireless channels from the communication and computation perspective.
As the study of the Metaverse is still in its infancy, only few works addressed the aspect of wireless resource allocation for the Metaverse. In~\cite{Han2021ADR}, an IoT-assisted Metaverse sync problem is studied in which an evolutionary game is formulated to enable the IoT devices to select VSPs to work for. However, the volume of the data and the limited bandwidth problem was not addressed. In contrast, our work mainly focuses on addressing the issue faced during data transmission from IoT devices to the VSP. Specifically, we propose the transmission of only semantic information to the VSP instead of raw data.
Since the semantic extraction from raw data depends heavily on the data type, we focus in this work on the case of camera and radar sensors for AV systems~\cite{Yang_2022, Ouaknine_2021_ICCV}. This is motivated by the fact that camera and radar sensors produce data with high volume and consume a large amount of bandwidth for transmission. However, our framework is straightforwardly to apply to other sensor types (e.g., voice recording) and other use cases of the Metaverse with changes only to the semantic valuation function.
We then leverage existing IoT devices such as existing AVs and smart phones to transmit their data to the VSP (Figure~\ref{fig:sys_model}). The VSP sends a UAV-BS to its region of interest to collect data.
Beside the cost efficiency, this solution is scalable as it enables the VSP to hire IoT devices to satisfy the targeted QoS.
However, IoT devices might take advantage of the importance of their semantic information to the VSP and increase their revenues by selling their data much higher than their true value. To overcome this manipulation of the strategic interaction, we develop a reverse auction mechanism that incentivizes IoT devices to make their bids to sell semantic information to the VSP truthfully. 
Our main contributions are as follows:
\begin{itemize}
    \item We propose a novel design for digital twin construction in the Metaverse by leveraging semantic information extraction algorithms. The obtained semantic information are sent to the VSP instead of the raw data, reducing the size of the transmitted data and consequently the latency.
    
    \item To collect data from IoT devices, we propose that the VSP sends a UAV-BS to its region of interest. A reverse auction mechanism is developed then to select the set of IoT devices that can access the wireless bandwidth and sell their semantic information to the VSP. 
    
    \item We conduct extensive simulations to validate our proposed model. The major finding is that the use of semantic information transmission enables more IoT devices to transmit their data to the VPS compared to traditional raw data transmission. This enables the VSP to have a variety of data sources and fast rendering of the digital twin in the Metaverse.
\end{itemize}

In Section~\ref{sec:system_moedl}, we describe our proposed semantic information marketing model. In Section~\ref{section:SW_max} we formulate the social welfare maximization problem. 
Simulation results are presented in Section~\ref{section:simulation} and Section~\ref{section:conclusion} concludes the paper.

\section{System Model: semantic information Market}\label{sec:system_moedl}

\subsection{Preliminaries About semantic information}
\begin{figure}[ht!]
    \centering
    \includegraphics[width=.40\textwidth,height=4.5cm]{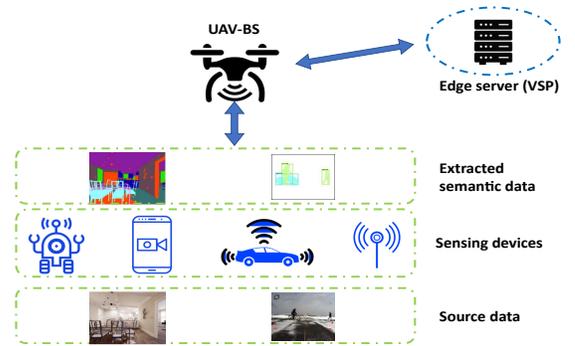}
    \caption{Exemplary use cases of semantic information for the Metaverse.}
    \label{fig:sys_model}
\end{figure}

As illustrated in Figure~\ref{fig:sys_model}, we consider a Metaverse market which consists of a virtual service provider (VSP) and a set of $\mathcal{N} = \{1, \dots, N\}$ IoT devices, i.e., data owners. The VSP is responsible for the construction of the digital twin in the Metaverse based on the collected data from the selected set of IoT devices. Each IoT device has a set of sensors to collect geo-spatial data from the surrounding environment and has a machine learning (ML) model to extract semantic information from the collected raw data. Different from traditional crowd-sensing platforms that collect all the raw data from the IoT devices, the VSP obtains only semantic information (e.g., semantic mask for each object with its corresponding class), which is motivated by the following reasons:
\begin{itemize}
    \item The number of channels available to the VSP are limited. Hence, if the VSP allows transmission of raw data by the IoT devices, only few devices will be able to transmit their data which might be of less importance than the data of other non-selected IoT devices.
    
    \item Raw data is large in size in general (e.g. video and images), which can increase the transmission delay making the rendering of the digital twin very slow and obsolete.
    
    \item The quality of the constructed digital twin will be higher as more semantic information about the physical world will be available to the VSP.
\end{itemize}

\subsection{Service Cost in the Metaverse Market}
The IoT device needs to calculate the cost of collecting data, extracting the semantics from raw data and the communication costs required to transmit the semantic information to the edge server. 

\subsubsection{Computation Cost}
The computation cost for data collection for IoT device $i$ is defined as follows~\cite{Yutao_2020_TMC}:
\begin{equation}\label{eq:c_i_R}
    c_i^R = \sum\limits_{k=1}^{m_i} d_{ik}^R \alpha_{ik}^R 
\end{equation}

\noindent where $d_{ik}^R$ is the size of the collected raw data and $\alpha_{ik}^R$ is the per unit computation cost for sensor $k$ by the IoT device $i$. $m_i$ is the number of sensors supported by the IoT device $i$. 
The cost of the semantic extraction is primarily based on the amount of energy required to perform the semantic extraction which can be formulated as
\begin{equation}
    c_i^{S} = \Delta t \gamma_i 
\end{equation}

\noindent where $\Delta t$ is the time required for the semantic extraction given the input raw data and $\gamma_i$ is the per time step computation cost.
The total computation cost is then
\begin{equation}\label{eq:c_i_p}
    c_i^p = c_i^R + c_i^{S}
     = \sum\limits_{k=1}^{m_i} d_{ik}^R \alpha_{ik}^R + \Delta t \gamma_i .
\end{equation}

\subsubsection{Communication Cost}

When transmitting semantic information to the VSP on $C_i$ channels with capacity $r$ for each channel\footnote{For simplicity, we consider a frequency-division multiple-access (FDMA) communication scheme, with other schemes straightforwardly applicable with appropriate adjustment to the cost function.}, the resulting communication cost to the IoT device $i$ is defined as~\cite{Yutao_2020_TMC}
\begin{equation}\label{eq:c_i_m}
    c_i^m = r C_i \beta_i 
\end{equation}

\noindent where $\beta_i$ is the per unit energy cost for transmission. 
By summing~\eqref{eq:c_i_p} and~\eqref{eq:c_i_m}, the total service cost is
\begin{equation}
    c_i = c_i^p + c_i^m
    = \sum\limits_{k=1}^{m_i} d_{ik}^R \alpha_{ik}^R +\Delta t \gamma_i + r C_i \beta_i .
\end{equation}

\subsubsection{Semantic Information Value}
Not all the extracted semantic information have the same value. For instance, if the IoT device has predicted a car collision, the device should value this information more and hence increase its price when selling the semantic information to the VSP. Therefore, the IoT device should include the valuation of the derived semantic information based on the sensitivity of its content in the final service cost valuation.
The semantic information value/reward is then defined as follows:
\begin{equation}\label{eq:R_i_S}
    R_i^{S} = \left[ \sum\limits_{j=1}^{O_t} \left( \delta^{j}_1 + \delta^{j}_2 + \delta^{j}_3 + \delta^{j}_4\right) q(j) \right] \frac{1}{w_t} + \frac{1}{d^R_i}
\end{equation}

\noindent where $O_t$ is the number of objects detected at time $t$, $\delta^{j}_1$, $\delta^{j}_2$, $\delta^{j}_3$ and $\delta^{j}_4$ are binary variable that reflects the existence of semantic information about the relative speed, size, relative position and moving direction for the object $j$, respectively. 
$q(j)$ is the model quality for the detected object $j$. For example, the quality function can be the intersection over unit (IoU) for image segmentation or the accuracy for object detection. Finally, $w_t$ is the weather condition at time $t$ scaled between zero and one. If the weather condition is bad, i.e., close to zero, the derived semantic information valuation $R_i^S$ will increase, reflecting its high value to the VSP.
The last term $\frac{1}{d^R_i}$ indicates the importance of data size to the VSP. For example, if two IoT devices have the same semantic value for certain data, the IoT device that provides this data with a smaller size is preferred by the VSP.

\subsection{Auction Based Metaverse Service Market}
In our considered system model, as there are several sellers (the IoT devices) and one buyer (the VSP), a reverse auction, where the traditional roles of buyer and sellers is reversed, is appropriate for our system design~\cite{Nisan_2007}. 
Therefore, to select the set of IoT devices to collect data from, the VSP conducts a reverse auction. 
Each IoT device bids for the price it is willing to sell its semantic information to the VSP. This is modeled as a competition on the channels provided by the VSP through its UAV-BS. Therefore, the IoT devices tries to increase their bids $\mathbf{b} =( b_{1}, b_{2}, \dots, b_{N})$ to gain access to the minimum number of channels required to transmit their data.
Since all sensors do not have the same data quality due to their location and hardware specifications, the VSP needs to choose the winners based on their semantic information quality in addition to data size and bid value.
The IoT device $i$ reports its type $t_i=\{b_i, R_i^S \}$ and the number of channels $C_i$ required to transmit its data. The IoT device is single minded, i.e., it either take all the request set of channels or any~\cite{Nisan_2007}.
Additionally, The VSP requires information freshness, i.e., the transmitted data should not exceed a certain time threshold $t_{max}$ which is broadcasted before the beginning of the auction. The IoT devices use this threshold to determine the required number of channels to transmit their local data within the specified threshold.




\section{Social Welfare Maximization Reverse Auction}\label{section:SW_max}
The IoT device's utility is the difference between its payment and service cost (computation cost and communication cost), which is expressed as
\begin{equation}\label{eq:u_i}
    u_i = p_i - R_i^{S} - b_i
\end{equation}

\noindent where $b_i=c_i$. The utility of the VSP is the sum of all IoT devices' semantic information rewards minus the cost $\hat{c}$ of allocating the channels from the wireless service provider and the sum of payments to the winners, which is written as
\begin{equation}
    \hat{u} = \sum\limits_{i=1}^{\mathcal{|N|}}\xi_i R_i^S  - \sum\limits_{i=1}^{\mathcal{|N|}}\xi_i p_i - \hat{c}
\end{equation}

\noindent where $\xi_i$ is a binary decision variable that indicates if IoT device $i$ is chosen among the winners ($\xi_i=1$) or not ($\xi_i=0$). The social welfare of the system is defined as the sum of the utilities of all the system entities (the VSP and the IoT devices), which is written as
\begin{equation}
    S(\xi) = \sum\limits_{i=1}^{\mathcal{|N|}}\xi_i u_i + \hat{u}
    = - \sum\limits_{i=1}^{\mathcal{|N|}} \xi_i b_i - \hat{c} .
\end{equation}

The social welfare is regarded as the system efficiency~\cite{Zhang_2017_welfare} and hence, maximizing the social welfare implies an efficient Metaverse market system. The social welfare maximization is formally written as an integer linear programming (ILP) problem as follows:
\begin{subequations}
\label{eq:optz_1}
\begin{align}
\begin{split}
\max_{\xi} S(\xi) = - \sum\limits_{i\in \mathcal{N}} \xi_i b_i - \hat{c}, \label{eq:MaxA} 
\end{split}\\
\begin{split}
\hspace{1cm} s.t. \sum_{i\in \mathcal{N}} \xi_i C_i \leq B \label{eq:MaxB}
\end{split}\\
\begin{split}
\hspace{1cm} \xi_i \in \{0,1\}, \forall i\in \mathcal{N} \label{eq:MaxC}
\end{split}
\end{align}
\end{subequations}

\noindent where $B$ is the number of channels provided by the VSP. The ILP presented in~\eqref{eq:optz_1} can be solved using a deterministic off-the-shelf solver or using a heuristic algorithm~\cite{Ismail_2021_Globecomm}. In Section~\ref{section:experiments}, we implement and compare both solutions.
To avoid manipulation of the market by malicious IoT devices, e.g., gaining a higher utility than deserved or getting a negative utility, the mechanism should guarantee the properties of incentive compatibility (IC) and individual rationality (IR).
Therefore, in the following, we present the payment rule for winning IoT devices and prove the properties of IC and IR.


    
    


\subsubsection{Payment Rule}
The payment rule for winning IoT devices is based on VCG mechanism payment~\cite{Zhang_2017_welfare} and is represented as follows:

\begin{equation}\label{eq:payment}
    p_k = S(\xi^*) - S_{\mathcal{N}\backslash\{k\}}(\varphi^*) + (R_i^{S} +b_k)\xi_k,
\end{equation}

\noindent where $\xi^*$ is the optimal allocation for IoT devices given the bidding and demand vectors, and $S(\xi^*)$ is the corresponding maximal social welfare obtained from~\eqref{eq:optz_1}. $S_{\mathcal{N}\backslash\{k\}}(\varphi^*)$ is the maximal social welfare when IoT device $k$ is not among the participant in the auction where $\varphi^*$ represents the corresponding optimal allocation strategy.

\subsubsection{Incentive Compatibility and Individual Rationality}
\begin{theorem}\label{them_1}
	The proposed VCG-based reverse auction mechanism is incentive compatible.
\end{theorem}
\begin{IEEEproof}
    We consider two cases for the IoT device $k$ in the set of bidders:\\
    \textit{Case 1:} The submitted bid $b_k$ by the IoT device $k$ is equal to its true valuation $c_k$. In this case, by substituting~\eqref{eq:payment} in~\eqref{eq:u_i}, the utility is written as
    \begin{equation}\label{eq:u_k_truthfull}
        u_k = S(\xi^*) - S_{\mathcal{N}\backslash\{k\}}(\varphi^*)
    \end{equation}
    
    \textit{Case 2:} The submitted bid $b_k^{'}$ by the IoT device $k$ is different from its true valuation, i.e., $b_k^{'} \neq c_k$ and the utility is written as
    \begin{equation}
        u_k^{'} = S(\xi^*{'}) - S_{\mathcal{N}\backslash\{k\}}(\varphi^*{'}) + \xi_k^*{'}b_k^{'} - \xi_k^*{'}c_k
    \end{equation}
    
    \noindent We note that the optimal allocation strategy in the absence of the IoT device $k$ in the auction in both cases is the same, i.e., $S_{\mathcal{N}\backslash\{k\}}(\varphi^*) = S_{\mathcal{N}\backslash\{k\}}(\varphi^*{'})$. The difference between the utilities of the above two cases is then calculated by:
    \begin{equation*}
         u_k^{'} - u_k = S(\xi^*{'}) - S(\xi^*) + \xi_k^*{'}b_k^{'} - \xi_k^*{'}c_k
    \end{equation*}
    \begin{equation*}
        = \left[ - \sum\limits_{i\neq k}^{\mathcal{|N|}} \xi_i^*{'} b_i - \xi_k^*{'} b_k^{'} \right] - 
        \left[ - \sum\limits_{i=1}^{\mathcal{|N|}} \xi_i^* b_i \right]
        + \xi_k^*{'}b_k^{'} - \xi_k^*{'} c_k
    \end{equation*}
    \begin{equation}
        = \left[ - \sum\limits_{i=1}^{\mathcal{|N|}} \xi_i^*{'} b_i \right] - 
        \left[ - \sum\limits_{i=1}^{\mathcal{|N|}} \xi_i^* b_i \right]
    \end{equation}
    
    If $u_k^{'} - u_k > 0$, then the IoT device has an incentive to deviate from $\xi^*$. However, since $\xi^*$ is the optimal solution for the ILP in~\eqref{eq:optz_1}, and hence better than any other solution (e.g., $\xi^*{'}$), we have $u_k^{'} \leq u_k$. Therefore, the IoT devices cannot obtain any benefit by misreporting their true valuation of their data. Note that the IoT device might submit a higher semantic value than its true value to gain access to the channels and sell its data with higher price. However, both the ILP problem in~\eqref{eq:optz_1} and the utility of the IoT device in~\eqref{eq:u_k_truthfull} are independent from the value of the submitted semantic value $R_i^{S}$, giving no incentive to the IoT device to submit untruthful value of $R_i^{S}$, concluding the proof.

\end{IEEEproof}

\begin{theorem}\label{them_2}
	The proposed VCG-based reverse auction mechanism has the property of individual rationality.
\end{theorem}
\begin{IEEEproof}
    Given that the mechanism is truthful, we consider two cases:
    
    \textit{Case 1:} $S(\xi^*) \geq S_{\mathcal{N}\backslash\{k\}}(\varphi^*)$, then based on~\eqref{eq:u_k_truthfull}, $u_k \geq 0$. Hence, the IoT device $k$ will get non-negative utility after joining the auction.
    
    \textit{Case 2:} $S(\xi^*) < S_{\mathcal{N}\backslash\{k\}}(\varphi^*)$. In this case, since $\mathcal{N}\backslash\{k\} \subset \mathcal{N}$, the VSP will have much more options to choose from with $\mathcal{N}$ rather than with $\mathcal{N}\backslash\{k\}$ implying that $S(\xi^*) \geq S_{\mathcal{N}\backslash\{k\}}(\varphi^*)$ and $u_k \geq 0$. Therefore, Case 2 cannot be realized and the proposed VCG-based reverse auction mechanism has the property of individual rationality.
\end{IEEEproof}

\section{Experiments}\label{section:experiments}\label{section:simulation}
In this section, we conduct experiments on real world data and present numerical results to evaluate the performance of our proposed reverse auction mechanism. Unless otherwise stated, we consider that $N=20$, $t_{max}=1s$, and that the channel's capacity for each IoT device is fixed to $r= 10 kbps$. 

\begin{figure}
\centering
\begin{subfigure}{.45\textwidth}
  \centering
  \includegraphics[width=.89\linewidth,height=1.75cm]{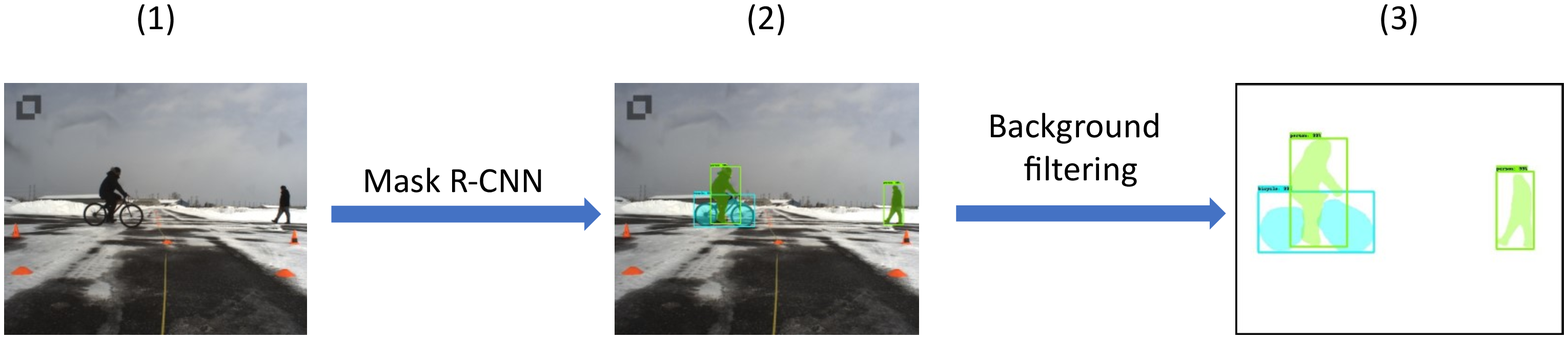}
  \caption{Camera view.}
  \label{fig:sem_camera}
\end{subfigure}\\
\begin{subfigure}{.45\textwidth}
  \centering
  \includegraphics[width=.89\linewidth,height=2.0cm]{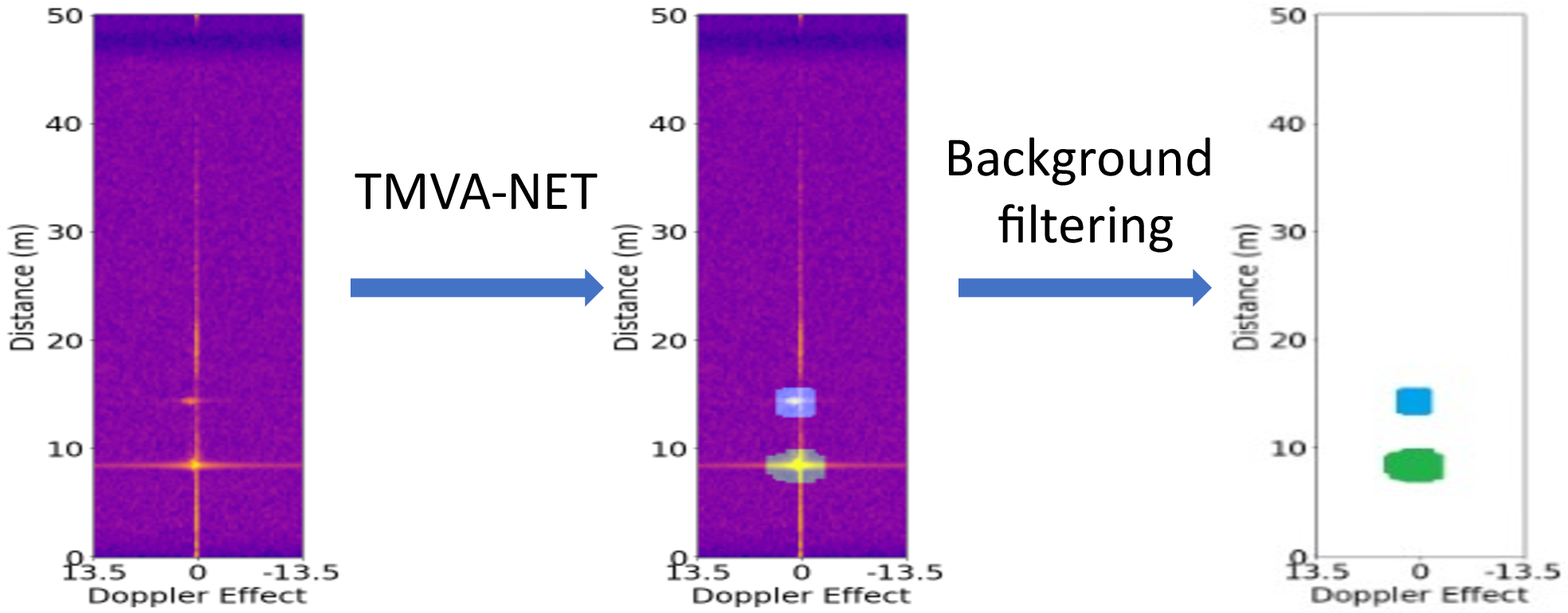}
  \caption{Range-Doppler view.}
  \label{fig:sem_RD}
\end{subfigure}\\
\caption{A scene from CARRADA dataset from different views with their respective semantic segmentation, with a cyclist and a pedestrian. The blue mask in (b) represents the pedestrian while the green mask represents the cyclist.}
\label{fig:sem_views}
\end{figure}

\subsection{Dataset and Experimentation's Settings}
We conduct our experiments on CARRADA dataset~\cite{Ouaknine_CARRADA_2020} which is a recent, open dataset that contains 30 scenes of synchronized sequences of camera and radar images. The radar images consists of the range-angle (RA) and the range-Doppler (RD) views. Figure~\ref{fig:sem_views} illustrates sample images from the CARRADA dataset where a cyclist and a pedestrian are moving into the opposite directions. To extract the semantic information from the raw data, different algorithms are used based on the supported sensors on the IoT device. We consider 4 types of IoT devices that use different semantic extraction algorithms. For devices with camera sensors only, Mask R-CNN algorithm for semantic segmentation and bounding boxes is used~\cite{He_2017_ICCV}. For devices with radar sensors only, RAMP-CNN (for RA view only), FCN-8s (for either RA or RD views) and, TMVA-NET (for RA and RD views) are used~\cite{Ouaknine_2021_ICCV, Ouaknine_CARRADA_2020, Gao_2020_RAMP_CNN}. 
We adopt simulation settings similar to~\cite{Ouaknine_2021_ICCV}. 
Each IoT device is considered to use one of the aforementioned algorithms based on a uniform distribution. 
For devices with both camera and radar sensors, a combination of these algorithms is used instead.
As an output for the aforementioned semantic segmentation algorithms, a white background image which contains the masked objects (see Figure~\ref{fig:sem_views}) is produced in addition to a meta-data text file (in JSON format) that contains a mapping between each object in the image with its class and other derived semantic information (e.g., range and shape). The meta-data text file has a maximum size of $d_{meta} = 1 kb$.









\subsection{Results}
\subsubsection{Impact of The Solver on The Social Welfare}
\begin{figure}[ht!]
    \centering
    \includegraphics[width=.40\textwidth,height=3.8cm]{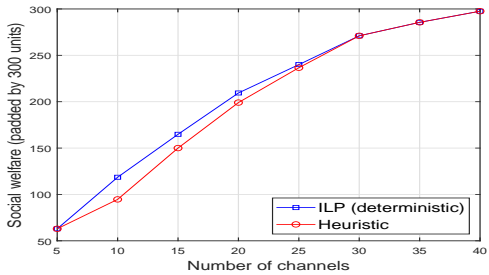}
    \caption{Impact of the solver on the social welfare.}
    \label{fig:exp_2}
\end{figure}

First, we evaluate the efficiency of our proposed ILP reverse auction algorithm. The ILP is solved deterministically using \emph{Gurobi optimizer} and the results are compared to the heuristic auction algorithm proposed in~\cite{Ismail_2021_Globecomm} with the payment rule adjusted to follow~\eqref{eq:payment}. 
We observe from Figure~\ref{fig:exp_2} that both solvers reach an equivalent optimal solution when the number of channels is higher than $30$. However, when the number of channels is lower, i.e., from $10$ to $25$, the heuristic algorithm is not able to reach the optimality derived by the deterministic optimizer.
This is explained by the fact that the heuristic algorithm first sorts the IoT devices (bidders) based on their bids and stops the search for potential winners once the social welfare stops increasing. This makes the algorithm deviate from the optimal solution as other potential combination of winners is not explored.
In what follows, we only consider the use of the deterministic optimizer to solve the social welfare maximization problem.

\subsubsection{Impact of semantic information Transmission on The Metaverse}
\begin{figure}[ht!]
    \centering
    \includegraphics[width=.40\textwidth,height=3.8cm]{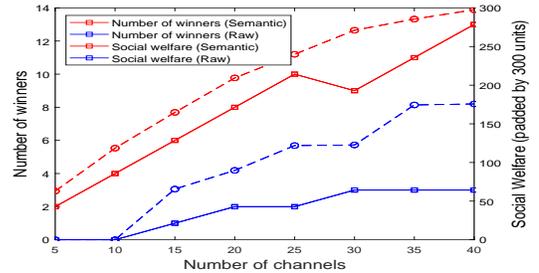}
    \caption{Impact of the number of channels on the number of winners and social welfare.}
    \label{fig:exp_1}
\end{figure}

We consider two types of transmission: the first one consists of transmitting raw data, i.e., no semantic information extraction, and the second one consists of transmitting only the semantic information extracted from the raw data. 
We observe from Figure~\ref{fig:exp_1} that as the number of channels $B$ provided by the VSP increases, the number of winners for the case of semantic information transmission also increases (up to 12 winners). However, in the case where only raw data is transmitted, the number of winners is very low ($1$ to $2$ winners) and almost remain unchanged, making the VSP relies heavily on the inputs of a small set of IoT devices. This result is explained by the fact that raw data has significantly a larger size compared to the size of semantic information and therefore one winner can use up to half of the available channels. Furthermore, even if the channels are fully allocated in both cases, the received data by the VSP might be insignificant in case of raw data transmission (e.g., in case of camera image, the captured image can be taken from an angle where no object is present). Specifically, if the VSP tries to create a copy of the physical world using the received raw data, the constructed virtual world would have a very poor quality and cannot capture the real dynamics of the system (e.g., congestion of the road). However, in case of semantic information transmission, the VSP has a variety of data sources/views which are rich of semantic information, e.g., instances and their shape or relative speeds, showing the importance of semantic information and heterogeneity of data sources. This result also indicates that our formulation of the utility functions is able to select IoT devices with more valuable semantic information as winners and hence, transmit their data to the VSP. 
An important point to highlight is that as more IoT devices send their data to the VSP, the chances of having redundant information increase, which helps creating a more reliable virtual environment.


Interestingly, we observe from Figure~\ref{fig:exp_1} that when the number of channels increases from $25$ to $30$, the number of winners drops by $1$. This is due to the fact that some IoT devices that have more valuable information but require a larger number of channels are not selected as winners when fewer channels are available, i.e., $B=25$. However, when the number of channels increases to $30$, the winner list changes to contain fewer IoT devices but they can provide a higher social welfare, which is clearly observable in Figure~\ref{fig:exp_1} (red dotted curve).

\subsubsection{Impact of the semantic information Extraction Algorithms and The Input Data on The Winner List}



\begin{figure}[ht!]
    \centering
    \includegraphics[width=.40\textwidth,height=3.8cm]{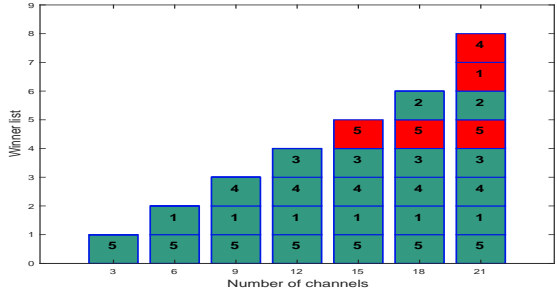}
    \caption{Impact of the extracted semantic information on the winner list. Green boxes represent IoT devices from group 1 and red boxes represent IoT devices from group 2.}
    \label{fig:exp_3}
\end{figure}

To observe how the winner list changes based on the semantic information extracted from data, we consider two identical groups of IoT devices where each group has 5 IoT devices and use different semantic information extraction algorithms as described in Table~\ref{table:1}. We provide the first group with scenes that contains 3 objects while the second group is provided with scenes which have only one object. The total number of participating IoT devices is then $N=10$ and the number of channels is varied from 3 to 21. We observe from Figure~\ref{fig:exp_3} that the first IoT devices to join the winner list after the number of available channels increases are from group 1. This is justified by the fact that input scenes for group 1 contains more semantic information than group 2. Interestingly, when the number of channels increases to 15, IoT device 5 from group 2 join the winner list leaving one element from group 1 not in the winner list. This is due to fact that IoT device 5 has camera and radar sensors and is using both Mask R-CNN and TMVA-NET algorithms for semantic extraction which have high quality of extracting significant semantic information. The remaining element from group 1 is IoT device 2 uses RAMP-CNN for radar semantic extraction, which has the lowest performance compared to other algorithms~\cite{Ouaknine_2021_ICCV}.

\begin{table}[ht!]
\begin{center}
\caption{Semantic extraction algorithms for different IoT devices.}
\begin{tabular}{ ||p{1.3cm}|p{2.20cm}|p{3.5cm}|| }
 \hline
  IoT device & Sensor(s) & Semantic algorithm(s) \\ 
 \hline\hline
  1 & camera & Mask R-CNN \\ 
 \hline
  2 & radar & RAMP-CNN (RA) \\ 
  \hline
  3 & radar & FCN-8s (RA)  \\
 \hline
  4 & radar & TMVA-NET (RA and RD)  \\ 
 \hline
  5 & camera \& radar &  Mask R-CNN + TMVA-NET \\ 
 \hline
\end{tabular}
\label{table:1}
\end{center}
\end{table}

\section{Conclusion}\label{section:conclusion}
In this paper, we presented a novel design to address the problem of data transmission over the wireless channels in the Metaverse. Specifically, we considered the scenario where a VSP is collecting data from the physical world using IoT devices in the field to create a digital twin of the road for AV training. We then proposed the use of semantic information extraction algorithms by the IoT devices to transmit only semantic information to the VSP instead of raw data. 
This significantly reduces the data transmission volume over the wireless network. 
It also enables more IoT devices to access the channels and the VSP to access a heterogeneity of data sources. As many IoT devices are interested in selling their semantic information to the VSP, we designed a reverse auction mechanism that has the properties of IC and IR to enable the VSP to choose the winners. Simulation results show that our proposed novel design is able to intelligently select IoT devices that have more semantic information and enable the VSP to create a high quality digital copy of the physical world. To further improve our design, we consider as a future work the use of a prediction model by the VSP to be trained on masked images to create scenes which have been lost during communication due to link failures.





\bibliographystyle{IEEEtran}
\bibliography{reference}

\end{document}